\documentclass[letterpaper]{aa}

\usepackage{txfonts} 
\usepackage[T1]{fontenc}
\usepackage{xspace}
\usepackage{graphicx}
\usepackage{natbib}

\newcommand{\arcsecsq}{\ensuremath{\mathrm{arcsec}^{2}}}
\newcommand{\kms}{\ensuremath{\mathrm{km\,s^{-1}}}}
\newcommand{\Lsun}{\ensuremath{L_{\sun}}}
\newcommand{\Msun}{\ensuremath{M_{\sun}}}
\newcommand{\Mbh}{\ensuremath{M_{\bullet}}}

\newcommand{\ML}[1][I]{\ensuremath{\Upsilon_{#1}}}
\newcommand{\MLI}{\ML[I]\xspace}

\newcommand{\hst}{{HST}\xspace}

\newcommand{\sauron}{\texttt{SAURON}\xspace}
\newcommand{\oasis}{\texttt{OASIS}\xspace}
\newcommand{\wfpcii}{\texttt{WFPC2}\xspace}
\newcommand{\fos}{\texttt{FOS}\xspace}
\newcommand{\pueo}{\texttt{PUEO}\xspace}

\newcommand{\objNGC}[1]{\object{NGC~#1}\xspace}
\newcommand{\objM}[1]{\object{M{#1}}\xspace}
\newcommand{\NGC}[1]{{NGC~#1}\xspace}

\graphicspath{{./}}
\DeclareGraphicsExtensions{{.eps}}

\newcommand{\fig}[1]{%
  \resizebox{\hsize}{!}{\includegraphics{#1}}
}

\newcommand{\figsc}[1]{%
  \sidecaption\includegraphics[width=12cm]{#1}
}

\newcommand{\ie}{\emph{i.e.,}\xspace}
\newcommand{\etal}{et~al.\xspace}
\newcommand{\schw}{Schwarzschild\xspace}

\newcommand{\reffig}[1]{Fig.~\ref{#1}}
\newcommand{\reftab}[1]{Table~\ref{#1}}
\newcommand{\refsec}[1]{Sect.~\ref{#1}}

\bibpunct{(}{)}{;}{a}{}{,} 

\begin{document}


\title{Axisymmetric dynamical models for \\ 
       \sauron and \oasis observations of \objNGC{3377}}

\author{Y.~Copin\inst{1,4}
\and    N.~Cretton\inst{2,5}
\and    E.~Emsellem\inst{3}
}
\institute{%
  Institut de Physique Nucléaire de Lyon, 69222 Villeurbanne, France
\and
  European Southern Observatory, Karl-Schwarzschild Strasse 2, 
  85748 Garching bei München, Germany 
\and 
  CRAL-Observatoire, 9 Avenue Charles-André, 69230 Saint-Genis-Laval, 
  France 
\and
  Sterrewacht Leiden, Postbus 9513, 2300 RA Leiden, The Netherlands 
\and
  Max-Planck Institut für Astronomie, Königstuhl 17, 
  69117 Heidelberg, Germany 
}

\titlerunning{Axisymmetric dynamical models of \NGC{3377}}

\offprints{Y. Copin, \email{y.copin@ipnl.in2p3.fr}}
\date{Received \ldots / Accepted \ldots}


\abstract{%
  We present a unique set of nested stellar kinematical maps of
  \NGC{3377} obtained with the integral-field spectrographs \oasis and
  \sauron. We then construct general axisymmetric dynamical models for
  this galaxy, based on the \schw numerical orbit superposition
  technique applied to these complementary measurements. We show how
  these two datasets constrain the mass of the central massive object
  and the overall mass-to-light ratio of the galaxy by probing the
  inner and outer regions respectively. The simultaneous use of both
  datasets leads us to confirm the presence of a massive black hole
  with a mass of $\Mbh = 7_{-5}^{+4}\,10^{7}$~\Msun{} (99.7\%
  confidence level), with a best-fit stellar mass-to-light ratio $\MLI
  = 2.1 \pm 0.2$ (for an assumed edge-on inclination).
}
\maketitle


\section{Introduction}

\NGC{3377} is a prototypical `disky' E5-6 galaxy with `boxy' outer
isophotes in the Leo~I group \citep[e.g.][]{1990AJ....100.1091P} at an
assumed distance of $D = 9.9$~Mpc. It has a power-law central
luminosity profile \citep{HST-centersIV, 2001AJ....121.2431R}, and its
total absolute magnitude of $\sim -19$ ($B$) is intermediate between
that of the classical `boxy' giant ellipticals and the `disky'
lower-luminosity objects \citep[e.g.][]{KB96}. Previous dynamical
models of this galaxy suggest the presence of a central massive black
hole (BH) of $\sim 10^8$~\Msun{} (\citealt[][hereafter
K+98]{ngc3377-K98}; \citealt[][hereafter G+03]{G3I-03}), while the
$\Mbh-\sigma$ relation \citep[and references
therein]{2002ApJ...574..740T} predict a BH of $\sim 4\,10^7$~\Msun.

In this paper, we present a unique combined set of nested
integral-field spectroscopic observations of \NGC{3377}, from
\sauron/William Herschel Telescope (WHT) and
\oasis/Canada-France-Hawaii Telescope (CFHT), on which we will base
our dynamical modeling.

\citet{sauron-paper1} presented the large-scale two-dimensional
stellar kinematics of \NGC{3377}, obtained with the panoramic
integral-field spectrograph \sauron as part of a representative survey
of nearby early-type galaxies \citep{sauron-paper2}. The resulting
kinematic maps cover $32\arcsec\times 43\arcsec$, with an effective
spatial resolution of $\sim 2\arcsec$~FWHM, and reveal modest but
significant deviations from axisymmetry, not only in the stellar
motions, but also in the morphology and kinematics of the
emission-line gas. This is unexpected, as lower-luminosity
steep-cusped systems such as \NGC{3377} were assumed to be
axisymmetric \citep[e.g.][but see
\citealt{2002ApJ...567..817H}]{1983ApJ...266...41D, HST-centersIII,
  Valluri98}.

Here we also present for the first time observations of the inner
$5\arcsec\times 6\arcsec$ of \NGC{3377} with a spatial resolution of
$\sim 0\farcs6$~FWHM, obtained at the CFHT with \oasis in its
adaptive-optics-assisted mode. The resulting nested set of
high-quality integral-field maps make \NGC{3377} a nearly ideal case
for detailed dynamical modeling, aimed at determining the stellar
mass-to-light ratio $M/L$, the internal orbital structure, and also
the BH mass \Mbh.

In this paper, we construct axisymmetric dynamical models which
incorporate all the integral-field kinematic data. Our approach is
similar to that followed by \citet{sauron-M32} for \objM{32}, but we use
an independent version of the modeling software, include all the
available integral-field datasets and study the full line-of-sight
velocity distributions (LOSVDs). By construction, we have to ignore
any signature of non-axisymmetry, but we will include them in triaxial
models in a follow-up work, in the way done by
\cite{2003astro.ph..1070V} for similar nested observations of
\objNGC{4365}. The comparison of the axisymmetric and triaxial models
will help clarify which (if any) of the parameters based on the
axisymmetric modeling are robust. This is relevant because many of the
nearly twenty galaxies for which BH masses have been determined by
means of (edge-on) axisymmetric models \citep{2002ApJ...574..740T,
  G3I-03} show similar signs of non-axisymmetry.

This paper is organized as follows. \refsec{sec:photometry} presents
the photometric data and the mass model adopted for \NGC{3377}, and
\refsec{sec:kinematics} describes the kinematic data. The dynamical
models are detailed in \refsec{sec:dynamical-models}, and discussed in
\refsec{sec:discussions}. \refsec{sec:concl} presents the conclusions.


\section{Photometry and mass model}
\label{sec:photometry}

In this Sect., we describe the construction of an accurate mass
model for \NGC{3377}, using the Multi-Gaussian Expansion (MGE) method
\citep{1992AA...253..366M,mge} applied to both wide-field and high
spatial resolution photometry. This allows us to probe the central
regions as well as to cover the full extent of the galaxy. The
detailed procedure is described in \citet[see also
\citealt{mge-m104,ngc4570-bar,CvdB99,mge-cappellari}]{ngc3115-dyn}.


\subsection{Photometric data}
\label{sec:phot-data}

\begin{figure*}
  \figsc{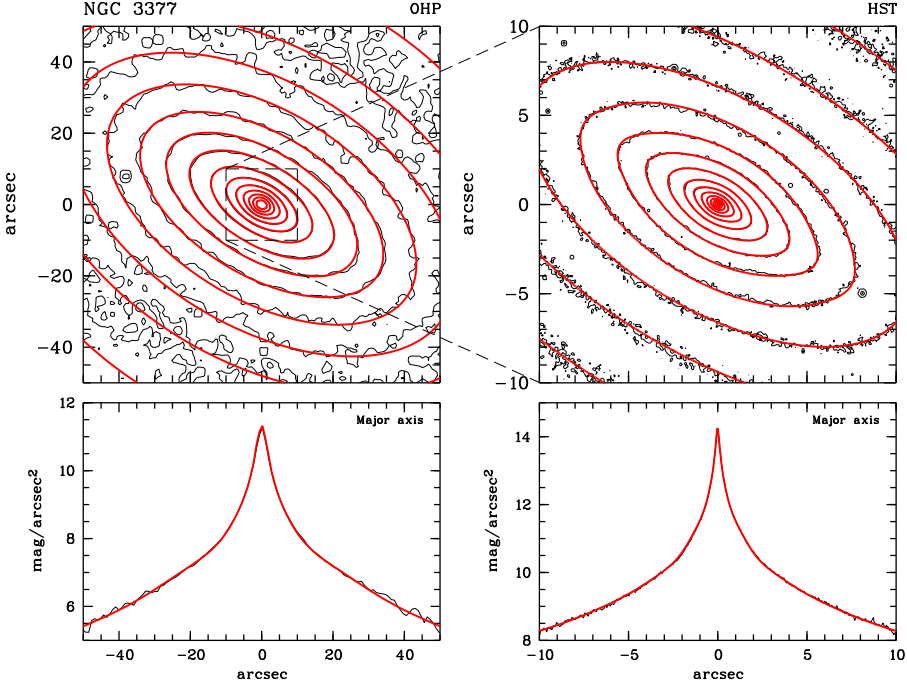}
  \caption{\emph{Upper panels:} $I$-band isophotes of \NGC{3377}
    (\emph{solid line}, step of 0.5~mag/\arcsecsq) and, superimposed,
    of the axisymmetric MGE model convolved with the appropriate PSF
    (\emph{heavy line}). \emph{Left:} wide-field image (OHP, courtesy
    of R.~Michard), \emph{right:} \hst/\wfpcii image. North is
    $20\degr$ (anti-clockwise) from the $y<0$-axis, and east is to the
    right. \emph{Lower panels:} cut along the major axis.}
  \label{fig:mge}
\end{figure*}

The large-scale image of \NGC{3377} shown in \reffig{fig:mge} (left
panel) was kindly provided by R.~Michard. It was obtained in 1995 with
the 1.2~m telescope of the Observatoire de Haute-Provence (OHP) in the
$I$-band, with a spatial resolution of $\sim 2\farcs1$~FWHM sampled at
0\farcs84, and a field-of-view (FoV) of $4\arcmin\times 7\arcmin$.
This image was reduced in the usual way (bias, dark, flat-field,
cosmic rays and cosmetics), and the sky contribution was estimated
from the outer part of the frame and then subtracted. The flux
normalization in \Lsun/pc$^{2}$ was carried out using published
aperture photometry \citep{poulain88,1994A&AS..104..179G}, accessible
via the HYPERCAT
service\footnote{\texttt{http://www-obs.univ-lyon1.fr/hypercat/}}.

For the high spatial resolution photometric data (see \reffig{fig:mge},
right panel), we use $I$-band (F814W) \hst/\wfpcii images,
retrieved through the ST/ECF archives (PI Faber, ID~5512). The
5~individual exposures ($2 \times 80$~s and $3 \times 350$~s) were
reduced in the standard way, and normalized in flux (in
\Lsun/pc$^{2}$) using the most recent PHOTFLAM conversion factors. The
exposures were combined (with cosmics removal) after verifying they
were properly centered.

The comparison between the two ground- and space-based flux calibrated
exposures shows an offset of $\sim 0.1$~mag, mostly due to the
difference in the zero points. We decided to use the \hst/\wfpcii
F814W image as a reference and renormalized the OHP~image accordingly.
The agreement between the two exposures is then excellent.


\subsection{The MGE surface brightness model}
\label{sec:mge}

We have used the Multi-Gaussian Expansion (MGE) method to model the
surface brightness distribution for \NGC{3377} using the photometric
data described above. The routine provides an analytical model of the
observed luminosity distribution, taking into account the convolution
effect of the point spread function (PSF). The method assumes that
both the PSF and the unconvolved surface brightness distribution of
the galaxy can be described by a sum of two-dimensional Gaussians,
whose best-fitting parameters are determined using an iterative
approach.

Each two-dimensional Gaussian $G'_{j}$ is described by its maximum
intensity $I'_{j}$, its dispersion $\sigma'_{j}$ along its major axis,
its flattening $q'_{j}$, its center given by its coordinates $X'_{j},
Y'_{j}$, and its position angle PA$_{j}$. In the general case, all
components are free to have different PA and centers, an $N$-Gaussian
model thus depending on $6N$ free parameters.

The MGE method has several advantages:
\begin{itemize}
\item the proper account of the individual PSFs allows the
  simultaneous use of complementary datasets with different spatial
  resolution (space- and ground-based observations);
\item its flexibility makes it well suited for complex,
  multi-component galaxies such as \NGC{3377}, which exhibits both
  diskyness and boxyness \citep[e.g.][]{1990AJ....100.1091P,
    1995A&A...293...20S};
\item the fitting procedure allows linear and non-linear constraints
  on the parameters. In the present case, since we require the
  deprojected model to be axisymmetric, we force the two-dimensional
  model to be bi-symmetric by imposing components to share the same PA
  and center, resulting in 3 free parameters per Gaussian.
\end{itemize}

The PSF of the OHP exposure was approximated with an MGE fit (two
concentric circular Gaussians) of the extracted image of a star
located several arc-minutes from the galaxy, where its background is
negligible. Using this model PSF, the large-scale surface brightness
distribution of \NGC{3377} is well described by a sum of 6~Gaussians.
Since the nuclear regions are strongly affected by seeing ($\sim
2\farcs1$~FWHM), we only keep the three outer components of this model
(with $14\farcs5 \leq \sigma'_{j} \leq 72\farcs4$, Gaussians $G'_{11}$
to $G'_{13}$ in the final model, see \reftab{tab:mge}), and proceed by
then fitting the central part using the \hst/\wfpcii exposure.

The PSF of the \hst/\wfpcii-\texttt{PC}/F814W exposure was derived
with the dedicated PSF simulator \texttt{TinyTim}%
\footnote{\texttt{http://www.stsci.edu/software/tinytim/tinytim.html}}
(v4.4), and consequently adjusted by a sum of three concentric
circular Gaussians. The presence of dust lanes in the very central
parts of \NGC{3377}, already noted by K+98, makes the determination of
the inner Gaussians more sensitive, and hence the convergence process
is slower. These dust lanes correspond to a maximal absorption of
$\sim 10\%$, and the areas affected were discarded from the MGE fit.
After subtraction of the properly aligned three-Gaussian `ground-based
model', the residual image was fitted by a sum of 10~Gaussians,
describing the inner parts of \NGC{3377} ($0\farcs04 \leq \sigma'_{j}
\leq 6\farcs04$, Gaussians $G'_{1}$ to $G'_{10}$ in the final model,
see \reftab{tab:mge}). Gaussians $G'_{5}$ and $G'_{8}$ are
significantly flatter than the others, betraying the presence of a
disk.

\begin{table}
  \caption{Parameters of the MGE deconvolved $I$-band surface
    brightness model of \NGC{3377}. $I'_{j}$, $\sigma'_{j}$ and
    $q'_{j}$ are the intensity, the dispersion along its major axis
    and the flattening, of the $j$-th two-dimensional Gaussian
    respectively (the flattest component is indicated in
    \emph{italics}). All the Gaussians share the same center and
    orientation. $I_{j}$ is the intensity of the three-dimensional 
    Gaussian resulting from the deprojection of the corresponding 
    two-dimensional Gaussian, assuming oblate geometry and an
    inclination of $i = 90\degr$ (see text).}
  \label{tab:mge}
  \begin{tabular}{llrrrr}
    \hline\hline
    Origin & Index & $I'_{j}$ & $\sigma'_{j}$ & $q'_{j}$ & $I_{j}$   \\
           & & [\Lsun/pc$^{2}$] & [\arcsec] & & [\Lsun/pc$^{2}$/\arcsec] \\
    \hline
      \hst & 1     & 536128.0 &  0.037        & 0.833    & 5833935.5 \\
           & 2     & 281535.8 &  0.105        & 0.753    & 1068021.9 \\
           & 3     & 107724.4 &  0.248        & 0.520    &  173261.0 \\
           & 4     &  53251.8 &  0.391        & 0.712    &   54347.3 \\
           & 5     &  12739.8 &  0.838        & 0.239    &    6066.5 \\
           & 6     &  30979.7 &  0.893        & 0.469    &   13838.3 \\
           & 7     &   9990.8 &  1.994        & 0.495    &    1998.9 \\
 & \emph{8} & \emph{2526.7} & \emph{3.598} & \emph{0.232} & \emph{280.2} \\
           & 9     &   2556.4 &  3.791        & 0.574    &     269.0 \\
           & 10    &   2337.2 &  5.989        & 0.479    &     155.7 \\
      OHP  & 11    &   1157.1 & 14.456        & 0.475    &      31.9 \\
           & 12    &    376.9 & 30.047        & 0.541    &       5.0 \\
           & 13    &     82.5 & 72.378        & 0.743    &       0.4 \\
    \hline
  \end{tabular}
\end{table}

The final MGE model of \NGC{3377} consists of the resulting set of
13~Gaussians (see \reftab{tab:mge}). \reffig{fig:mge} displays the
contour maps of the $I$-band images with superimposed contours of the
appropriately convolved MGE model, as well as cuts along the major
axis. The resulting fit to the observed photometry is excellent. There
is no sign of departures from axisymmetry in the central regions, and
the PA of the major axis is constant with radius
\citep[e.g.][]{1990AJ....100.1091P}. The total luminosity of the MGE
model is $L_{I} = 1.1\,10^{10}$~\Lsun, yielding $M_{I} = -21.0$.
Taking a total $B-I = 1.85$ \citep{Idiart02}, we get $M_{B} = -19.15$,
in good agreement with the value of $M_{B} = -19.2$ available in the
LEDA database%
\footnote{\texttt{http://leda.univ-lyon1.fr}} \citep{ledaI}.


\subsection{Deprojection}
\label{sec:deproj}

The MGE modeling allows the spatial luminosity density to be
analytically derived from the surface brightness model, assuming an
inclination angle $i$, and that the luminosity density associated with
each individual three-dimensional Gaussian is stratified on concentric
ellipsoids. Each two-dimensional Gaussian $j$ then uniquely deprojects
into a three-dimensional Gaussian, whose parameters $I_{j}$,
$\sigma_{j}$ and $q_{j}$ can be derived from $I'_{j}$, $\sigma'_{j}$,
$q'_{j}$ and $i$ following relations provided in
\citet{1992AA...253..366M}. The associated gravitational potential
$\Phi$ can be derived easily assuming a given (constant) mass-to-light
ratio \MLI \citep[see][for details]{mge}.

The inclination $i$ of the best-fitting MGE model (see \reftab{tab:mge})
is (formally) constrained by that of the flattest two-dimensional
Gaussian to be $90\degr \geq i \geq 76.6\degr =
\arccos{(\min\{q'_{j}\} = 0.232)}$. This constraint should however be
treated with some caution since it is model dependent. However, no
good MGE fit could be found with the additional requirement
$\min\{q'_{j}\} \geq 0.35$. This sets a robust minimum inclination of
$i_{\min} \simeq 70\degr$. In the following, we will only consider the
edge-on model, \ie $i = 90\degr$, since the inclination hypothesis
is likely not to be the most restrictive one (see discussion in
\refsec{sec:discussions}).


\section{Kinematical data}
\label{sec:kinematics}

In this Sect., we summarize the stellar kinematic measurements
obtained with the integral-field spectrographs \oasis and \sauron.


\subsection{\oasis observations}
\label{sec:oasis-obs}

\NGC{3377} was observed on April~1 and~2 1998 with \oasis mounted on
the adaptive optics (AO) bonnette \pueo of the Canada-France-Hawaii
Telescope \citep[see][for a full description of
\oasis{}]{oasis98,m31-oasis}. In order to take advantage of the better
performance of adaptive optics in the red, we used the MR3
configuration covering the \element{Ca} triplet ($\sim 8500$~\AA)
region. We selected a spatial sampling of 0\farcs16 per
hexagonal-shape lens, which provides a FoV of $6\farcs2 \times
5\farcs0$. The instrumental setup is given in
\reftab{tab:oasis-config}.

\begin{table}
  \caption{Instrumental setup for the \oasis observations.}
  \label{tab:oasis-config}
  \begin{tabular}{ll}
    \hline\hline
    \multicolumn{2}{c}{\pueo}  \\
    \hline
    Loop mode     & automatic  \\
    Loop gain     & 80         \\
    Beam splitter & I          \\
    \hline
    \multicolumn{2}{c}{\oasis} \\
    \hline
    Spatial sampling  & 0\farcs16                  \\
    Field-of-view     & $6\farcs2 \times 5\farcs0$ \\
    Number of spectra & 1123                       \\
    Spectral sampling & 2.23 \AA~pixel$^{-1}$      \\
    Spectral resolution ($\sigma$) & 70~\kms       \\
    Wavelength range  & 8351--9147 \AA             \\
    \hline
  \end{tabular}
\end{table}

Seven 30~mn exposures centered on the nucleus (but slightly dithered
to avoid systematics) were acquired, the AO being locked on the
central cusp of the galaxy. The atmospheric conditions were
photometric, and observations were carried out at low airmass
($<1.16$). Natural seeing conditions were mediocre with FWHM between
1\arcsec{} and 1\farcs5, providing an AO corrected PSF with FWHM~$\sim
0\farcs6$ (see \refsec{sec:psf-merge}). Neon arc lamp exposures were
obtained before and after each object integration. Other
configurations exposures (bias, dome flat-field, micro-pupil) were
usually acquired at the beginning or the end of the nights. Twilight
sky exposures were obtained at dawn or sunrise.


\subsection{\sauron observations}
\label{sec:sauron-obs}

\NGC{3377} was observed on February~17 1999 with \sauron during its
first scientific run on the William Herschel Telescope
\citep{sauron-paper1}. The instrumental setup is summarized in
\reftab{tab:sauron-config}. Four centered but slightly dithered 30~mn
exposures were acquired, providing a total FoV of
$32\arcsec\times 43\arcsec$ with an original sampling of 0\farcs94. The
spectral coverage of \sauron includes the \element{Mg} absorption
triplet as well as \element{Fe} and \element{Ca} absorption lines, and
the [\ion{O}{iii}] and \element{H}$\beta$ emission lines when present.

\begin{table}
  \caption{Instrumental setup for the \sauron observations.}
  \label{tab:sauron-config}
  \begin{tabular}{ll}
    \hline\hline
    \multicolumn{2}{c}{\sauron} \\
    \hline
    Spatial sampling  & 0\farcs94                \\
    Field-of-view     & $33\arcsec\times 41\arcsec$ \\
    Number of spectra & 1431                     \\
    Spectral sampling & 1.1~\AA~pixel$^{-1}$     \\
    Spectral resolution ($\sigma$) & 108~\kms    \\
    Wavelength range  & 4820--5340 \AA           \\
    \hline
  \end{tabular}
\end{table}


\subsection{IFS data reduction}
\label{sec:data-red}

\subsubsection{Spectra extraction and calibration}
\label{sec:extract-calib}

The integral-field spectroscopic data were reduced according to the
usual procedure described in \citet{m31-oasis} for \oasis and
\citet{sauron-paper1} for \sauron. The standard reduction procedures
include CCD preprocessing, optimal extraction of the spectra,
wavelength calibration, spectro-spatial flat-fielding and cosmic-ray
removal.

Given the small FoV of \oasis and the high surface brightness of the
nucleus of \NGC{3377} ($\mu_{V} = 17.7$ at 3\farcs3), no sky
subtraction was required for this instrument. Furthermore, no flux
calibration was performed for \oasis, since this is not required for
measurement of the stellar kinematics.

On the \sauron exposures, after spectral resolution rectification to a
common value of 1.91~\AA~$\sim 108$~\kms \citep[see][for
details]{sauron-paper1}, the night-sky spectrum was estimated from the
dedicated sky lenslets and subtracted from the object spectra.

\subsubsection{Spatial PSF estimates and datacubes merging}
\label{sec:psf-merge}

The knowledge of the spatial PSF is of particular importance in the
modeling process, as we aim at using datasets with very different
spatial resolutions. The spatial PSFs presented below have been
estimated from the comparison between the images reconstructed from
our IFS data and higher resolution reference frames \citep[see details
of the method in][]{m31-oasis}. Once the individual frames were
properly aligned and renormalized, the merging of multiple exposures
was carried out according to the prescriptions detailed in
\citet{sauron-paper1}. Note that the data reduction software
simultaneously provides the variance along each spectrum, which will
used to estimate the local signal-to-noise in the datacube.

\begin{figure}
  \fig{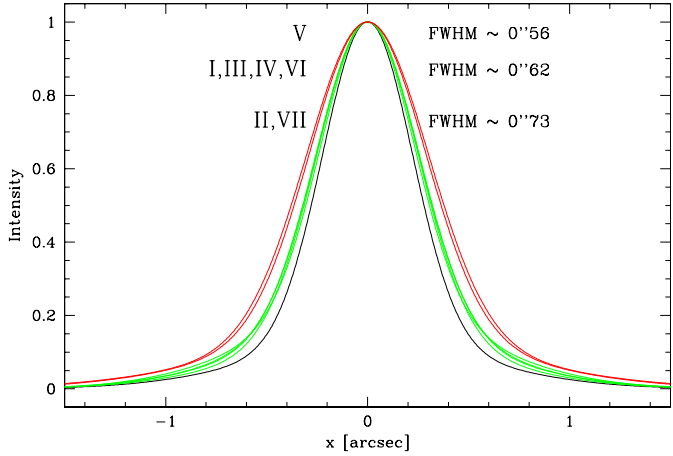}
  \caption{Fitted PSF of the 7~\oasis individual exposures. The PSFs
    have been approximated by a sum of 2~concentric circular Gaussians
    (see text). The individual PSFs can be sorted in 3~categories,
    with a mean FWHM of 0\farcs56, 0\farcs62 and 0\farcs73
    respectively.}
  \label{fig:oasis-psf}
\end{figure}

\paragraph{\oasis.}
The PSF of the individual \oasis exposures is well approximated by a
sum of two concentric circular Gaussians, and we estimated it by
comparison with the \hst/\wfpcii~F814W image described in
\refsec{sec:phot-data}. We ranked the 7~\oasis PSFs, labeled from
\textsc{i} to \textsc{vii}, according to their FWHM (see
\reffig{fig:oasis-psf}): 0\farcs56 for exposure \textsc{v}, $\sim
0\farcs62$ for exposures \textsc{i},\textsc{iii},\textsc{iv} and
\textsc{vi}, and $\sim 0\farcs73$ for exposures \textsc{ii} and
\textsc{vii}.

\begin{figure}
  \fig{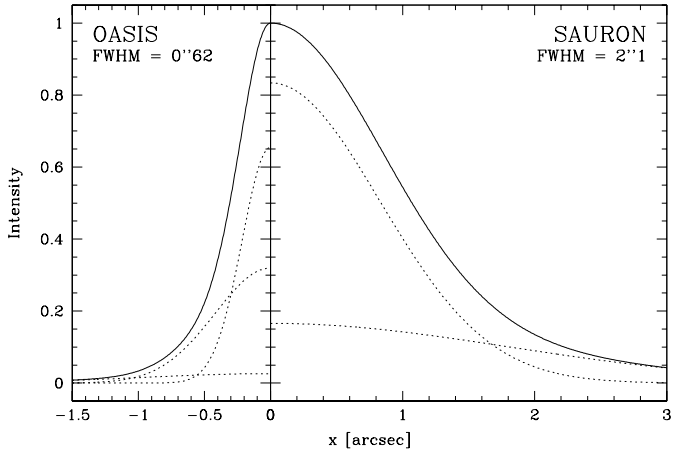}
  \caption{Effective PSF for the \oasis (\emph{left}) and
    \sauron (\emph{right}) merged datacubes of \NGC{3377}. As
    described in the text, the \oasis PSF is highly non-Gaussian and
    is properly described by a sum of three Gaussians (\emph{dotted
      lines}); the spatial resolution is estimated to be
    0\farcs62~FWHM. The \sauron PSF is described by a double
    Gaussian of 2\farcs1~FWHM.}
  \label{fig:3377-psf}
\end{figure}

Single exposures have a signal-to-noise ratio $S/N$ which is
insufficient for individual use in deriving the stellar kinematics. We
have therefore constructed two merged datacubes using two different
sets of exposures (trading-off between $S/N$ and resolution):
\begin{description}
\item[Datacube `A':] includes all the individual exposures except
  cubes \textsc{ii} and \textsc{vii}, in order to optimize the spatial
  resolution;
\item[Datacube `B':] includes all seven individual exposures, in order
  to maximize $S/N$.
\end{description}
We approximate the PSF of each merged datacube by a sum of three
Gaussians (see \reftab{tab:psf} and \reffig{fig:3377-psf}). It appears
that there is not much difference in terms of effective spatial
resolution between cubes `A' and `B', while the global $S/N$ of
datacube `B' is slightly higher. Hence we consider only cube `B',
which consists of 637~spectra with a final sampling of 0\farcs25.

\begin{table}
  \caption{PSF parameters of the \oasis and \sauron merged exposures.}
  \begin{tabular}{lcccccc}
  \hline\hline
  ID & FWHM & 
      $\sigma_1$ & $\sigma_2$ & $I_2/I_1$ & $\sigma_3$ & $I_3/I_1$ \\
  \hline
  \oasis/A & 0\farcs61 & 
      0\farcs22  & 0\farcs45  & 0.33      & 1\farcs11  & 0.02 \\
  \oasis/B & 0\farcs62 & 
      0\farcs21  & 0\farcs42  & 0.49      & 0\farcs96  & 0.04 \\
  \hline
  \sauron & 2\farcs15 & 
      0\farcs83  & 1\farcs80  & 0.20 & & \\
  \hline
  \end{tabular}
  \label{tab:psf}
\end{table}

\paragraph{\sauron.}
All four individual \sauron exposures have similar spatial resolution,
with a FWHM ranging from 2\farcs0 to 2\farcs5 (estimated from direct
images obtained before and after the \sauron observations). All of
them were therefore merged in a cube of 2957~spectra over-sampled at
0\farcs68 to allow a proper analysis of the spatial PSF.

We approximate the spatial PSF of the \sauron merged datacube by a
sum of two Gaussians and estimated their parameters by comparison with
the \hst/\wfpcii~F555W images obtained and reduced as described in
\refsec{sec:phot-data} (see \reftab{tab:psf} and
\reffig{fig:3377-psf}). The \sauron PSF is significantly non-gaussian:
the two-Gaussian approximation has a global FWHM of 2\farcs15, while
the best fit with a single Gaussian has a FWHM of 2\farcs62.

In order to reduce significantly the amount of data to handle during
the modeling, while retaining all the initial information, the four
individual \sauron data-cubes were then merged again in a final cube
of 1534~spectra sampled at 0\farcs94, corresponding to a strict
Nyquist sampling of the spatial PSF.

\subsubsection{Spatial binning}
\label{sec:binning}

In order to increase the $S/N$, and reduce the number of independent
apertures for the dynamical modeling, we applied an adaptive quadtree
spatial binning to the \sauron datacube (see
Appendix~\ref{sec:2d-binning}). We did not apply this technique to the
\oasis datacube, as we wished to retain the highest spatial resolution
available. The \sauron binned datacube finally includes
475~independent spectra (see \reffig{fig:binning}).


\subsection{Stellar kinematics}
\label{sec:kin-stellar}

We obtained reference stellar templates from dedicated exposures of
HD~132737 for \oasis, and HD~85990 for \sauron, both of spectral type
K0III. The stellar spectra were optimally summed over a spatial
aperture of $R = 2\arcsec$ to maximize the $S/N$ ($\sim 1300$ for
\oasis and $\sim 500$ for \sauron). The resulting individual stellar
spectra, as well as the final \oasis and \sauron \NGC{3377} datacubes,
were then continuum divided and rebinned in $\ln\lambda$. We used our
own version of the Fourier Correlation Quotient (FCQ) method
\citep{FCQ,BSG94} to derive the non-parametric LOSVD at every point of
the final datacubes. When needed, the LOSVDs were parametrized using a
simple Gaussian and complemented using 3rd and 4th order Gauss-Hermite
moments $h_{3}$ and $h_{4}$ \citep{vdMF93, Gerhard93}. We checked that
similar results were obtained with an updated version of the
cross-correlation method \citep{CCF}, and that these did not
significantly depend on the details of the continuum subtraction
procedure.

\reffig{fig:oasis-kin} and \ref{fig:sauron-kin} present the kinematic
maps (mean velocity $V$, velocity dispersion $\sigma$, Gauss-Hermite
moments $h_{3}$ and $h_{4}$) for \oasis and \sauron respectively, in
the setup used for the dynamical modeling (see
\refsec{sec:dynamical-models}).

\begin{figure*}
  \fig{0076-f04}
  \caption{\oasis stellar kinematical maps (\emph{filtered}:
    $S/N_{\min} = 40$). In each panel, the hatched \emph{disk}
    corresponds to the size of the seeing disk of 0\farcs62~FWHM.}
  \label{fig:oasis-kin}
\end{figure*}

\begin{figure*}
  \fig{0076-f05}
  \caption{\sauron stellar kinematical maps. In each panel, the
    hatched \emph{disk} corresponds to the size of the seeing disk
    of 2\farcs1~FWHM.}
  \label{fig:sauron-kin}
\end{figure*}

Estimates of the errors on each velocity bin of the non-parametric
LOSVDs are obtained by means of a Monte-Carlo approach
\citep[see][]{PhD-copin}: for each galaxy position, a noise-free
galaxy spectrum is built by convolving the template spectrum with the
corresponding parametrized LOSVD. We then add noise realizations
(consistent with the derived noise spectra of the datacube), and
extract the LOSVD via FCQ. The error is defined as the variance of the
distribution at each velocity bin after 100~realizations. If needed,
errors on the measured kinematical parameters are derived from the
same procedure, which produces reasonably realistic error-bars (see
e.g. \reffig{fig:OvsS-psf} and \ref{fig:kin-comp}; see also
\citealt{2003MNRAS.339..215D}).

\subsubsection{Comparisons with different datasets}
\label{sec:3377-comp}

\begin{figure}
  \fig{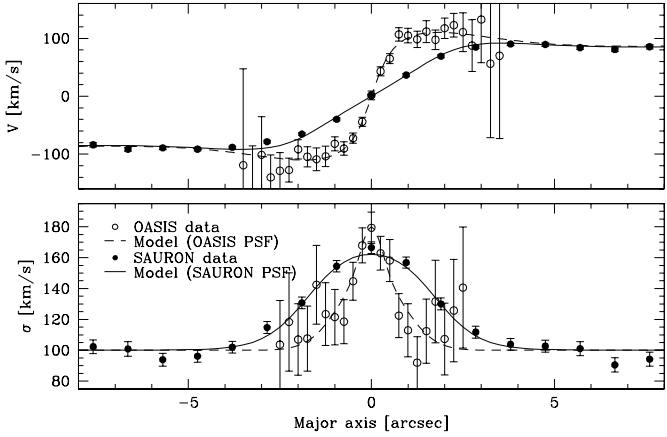}
  \caption{Comparison along the major axis between the \oasis
    (\emph{open circles}) and \sauron (\emph{closed circles}) datasets
    (\emph{upper panel:} mean velocity, \emph{lower panel:} velocity
    dispersion), and the simple analytic models convolved to the
    proper \oasis (\emph{dashed line}) and \sauron (\emph{solid line})
    resolutions.}
  \label{fig:OvsS-psf}
\end{figure}

We checked that the \oasis and \sauron datasets were consistent with
each other, not only regarding their surface brightness distribution
from which the PSFs were estimated, but also by comparing $V$ and
$\sigma$ maps. The \oasis FoV is too small to allow proper convolution
to the \sauron resolution. We therefore built a simple analytical
model which approximates the `intrinsic' $V$ and $\sigma$ fields, and
used the MGE photometry model to convolve this model with the adequate
PSFs. The agreement between the kinematics obtained with \oasis and
\sauron is excellent, as shown in \reffig{fig:OvsS-psf}.

\begin{figure*}
  \fig{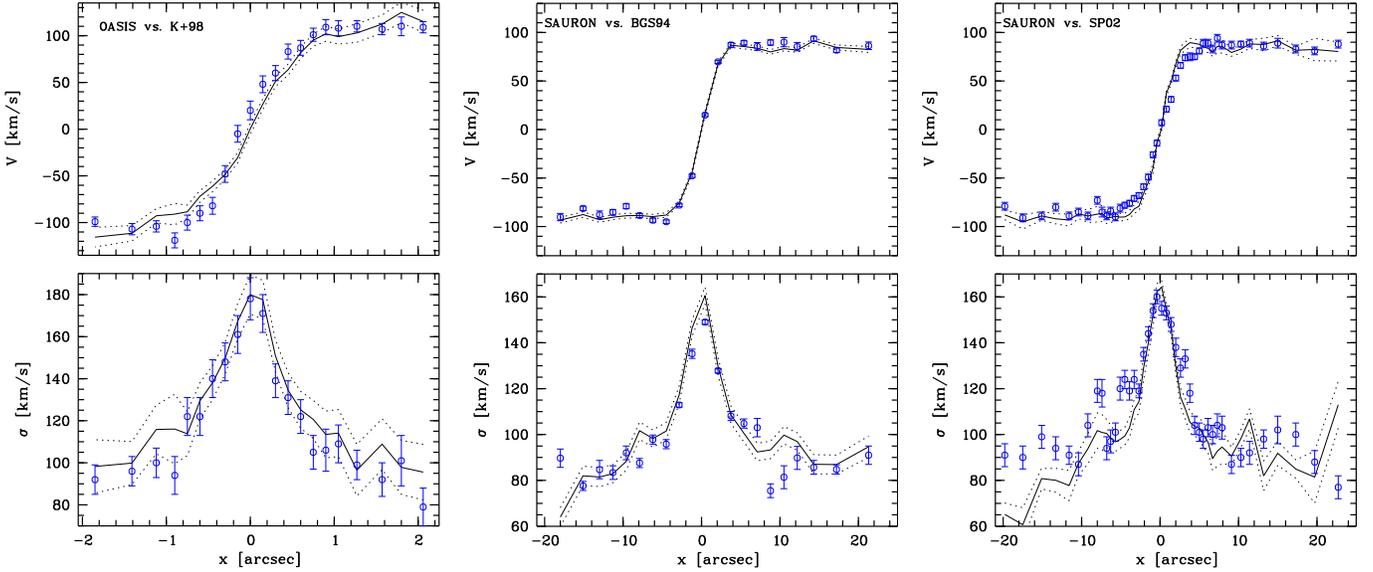}
  \caption{Comparison between the long-slit stellar kinematics from 
    literature (\emph{points}) and the equivalent IFS kinematics
    (\emph{lines}) for the mean velocities (\emph{top}) and velocity
    dispersions (\emph{bottom}): \oasis vs. K+98 (\emph{left}),
    \sauron vs. BSG94 (\emph{middle}) and \sauron vs. SP02
    (\emph{right}). The \emph{dotted lines} give the 1-$\sigma$ error
    on the IFS parameters, as estimated from Monte-Carlo simulations.}
  \label{fig:kin-comp}
\end{figure*}

We also compared the \oasis and \sauron stellar kinematics with
measurements based on long-slit data published by \citet[Calar Alto,
slit width of 2\farcs1, spectral resolution of 46~\kms, hereafter
BSG94]{BSG94}, K+98 (spectrograph \texttt{SIS}/CFHT, slit width $\sim
0\farcs3$--0\farcs5, spectral resolution $\sim 40$--60~\kms) and
\citet[spectrograph \texttt{CARELEC}/OHP, slit width of 1\farcs5,
spectral resolution of 25~\kms, hereafter SP02]{2002A&A...384..371S}.
A proper comparison between these different datasets is difficult
because of the very different spatial resolutions and instrumental
setups. We therefore restricted these comparisons to the datasets
which share \emph{similar} spatial resolutions, \ie \oasis with K+98
and \sauron with BSG94 and SP02. The IFS data were thus binned
according to the characteristics of the long-slit data and compared as
shown in \reffig{fig:kin-comp}: the datasets are in good agreement,
particularly considering the difficulty mentioned above. The
(marginal) 0\farcs15-offset reported by K+98 in the mean velocity
profile is not seen in our \oasis data.

We chose not to use the \fos observations of the nucleus of \NGC{3377}
(G+03, square aperture of 0\farcs21, spectral resolution of $\sim
100$~\kms, no aperture illumination corrections applied) for three
reasons: (a) the lack of extensive spatial information makes the
comparison with such high-spatial resolution data hazardous, (b) the
proper determination of aperture positioning is known to be a critical
procedure \citep{m32-fos}, potentially leading to critical
consequences for the dynamical modeling, (c) by excluding these
measurements, our dynamical models are \emph{fully} independant to the
ones developed by G+03.

\subsubsection{First results from the \sauron and \oasis maps}
\label{sec:IFSmaps}

As mentioned earlier, the \oasis and \sauron kinematics presented in
\reffig{fig:oasis-kin} and \ref{fig:sauron-kin} respectively are fully
coherent and in excellent agreement with the long-slit datasets
previously published, up to the highest spatial resolutions. The steep
velocity gradient, unresolved at the \sauron resolution, reaches a
maximum of $\sim 110$~\kms{} at $r \approx 1\farcs5$, and then decreases
to a plateau of $\sim 90$~\kms. The velocity dispersion peak is
barely spatially resolved in the \oasis observations, and reaches $\sim
170$~\kms. The $h_{3}$ coefficient is anti-correlated with the mean
velocity as expected from the superposition of a bright dynamically
cold component on a hotter spheroid.

Integral-field spectroscopy enables us to analyse the `morphology' of
the kinematics, something not possible using long-slit spectra. Both
\oasis and \sauron mean velocity fields reveal a twist of the zero
velocity curve of $\sim 10^{\circ}$ with respect to the photometric
minor-axis up to $r \approx 10\arcsec$, while the photometry does not
show any variation of position angle \citep{1990AJ....100.1091P}. As a
consequence, the velocities along the minor-axis are not null, a fact
not reported so far\footnote{Long-slit (unpublished) data showed a
  weak rotation along the minor-axis of \NGC{3377} \citetext{Tom Statler,
  private communication}.}, and which is clearly visible with an
amplitude of $\sim 7$~\kms{} for \oasis and $\sim 12$~\kms{} for
\sauron.

The \sauron velocity dispersion-map also presents interesting
morphological features: while the $\sigma$-peak is elongated along the
photometric minor-axis --~consistent with the diskyness of the light
distribution~--, it displays a twist of about $10^{\circ}$ in the
direction \emph{opposite} to the kinematic misalignment. These modest
but significant departures from axisymmetry are probable signatures of
a triaxial intrinsic shape. This is also apparent in the ionized gas
distribution and motions, which exhibit a noticeable spiral-like
morphology and strong departures from circular motions
\citep{sauron-paper1}. These issues will be discussed in more detail
in a forthcoming paper \citetext{Emsellem \etal, in preparation}.


\section{Dynamical models}
\label{sec:dynamical-models}


\subsection{\schw models}
\label{sec:schw}

We construct dynamical models based on the orbit superposition
technique of \citet{Sch79,Sch82}. Similar orbit-based models were
constructed to measure the masses of central BHs \citep{m32-axisym,
  CvdB99, sauron-M32, 2002ApJ...578..787C, G3I-03}, and dark halo
parameters \citep{rix97, 1998MNRAS.295..197G, 2000AJ....119..153S,
  2000A&AS..144...53K}. Our implementation is described in detail in
\citet[hereafter C99]{cretton99}. We summarize it briefly here.

We sample the stellar orbits using a grid in integral space, which
includes: the energy $E$, the vertical component of the angular
momentum $L_{z}$ and an effective third integral $I_3$. We use 20~values
of $E = 1/2\; R_c \; \partial \Phi/\partial R + \Phi(R_c,0)$, sampled
through the radius of the circular orbit $R_c$. We take a logarithmic
sampling of $R_c$ in $[0\farcs01,300\farcs0]$, since more than
99\% of the total mass of our MGE model lies inside this range.
14~values of $L_z$ per $E$ in $[-L_{z,\max},+L_{z,\max}]$ and 7~values
of $I_3$ per ($E,L_z$) were adopted (see C99).

The orbit library is constructed by numerical integration of each
trajectory for an adopted amount of time (200~periods of the circular
orbit at that $E$) using a Runge-Kutta scheme. During integration, we
store the fractional time spent by each orbit in a Cartesian
`data-cube' $(x', y', v_{\mathrm{los}})$, where $(x', y')$ are the
projected coordinates on the sky and $v_{\mathrm{los}}$, the
line-of-sight velocity. We use an $E$-dithering scheme (see C99) to
make each orbit smoother in phase space. The data-cube of each
individual orbit is further convolved with the PSF and eventually
yields the \emph{orbital} LOSVD $\mathcal{L}_{(x',y')}
(v_{\mathrm{los}})$ at each position $(x',y')$ on the sky.

In C99, we adopted a parametrized form for the (orbital and observed)
LOSVDs using the Gauss-Hermite series expansion. However, as
emphasized in \citet{CvdB99}, some problems may arise in the modeling
of dynamically `cold' systems (\ie with a high $V/\sigma$), where the
use of the linear expansion can lead to spurious counter-rotation. To
avoid this potential shortcoming, we choose to constrain the dynamical
models using directly the full non-parametrized LOSVDs. As a
consequence, the maps presented in \reffig{fig:oasis-kin} and
\ref{fig:sauron-kin} are shown for illustration purposes, but were not
used in the modeling technique.

Orbital occupation times are also stored in logarithmic polar grids in
the meridional plane and in the $(x',y')$ plane, to make sure the
final orbit model reproduces the MGE mass model. The mass on each
orbit is computed with the NNLS algorithm \citep{nnls}, such that the
non-negative superposition of all orbital LOSVDs aims at reproducing
the observed LOSVDs within the errors. In addition, the model has to
fit the intrinsic and projected MGE mass profiles. Smoothness in
integral space can be enforced through a regularization technique (see
C99). Models with different values of BH mass \Mbh{} and mass-to-light
ratio \MLI are constructed and compared to the data. As mentioned in
\refsec{sec:deproj}, the mass-to-light ratio \MLI is assumed to be
constant over the whole extent of the galaxy. The quality of the fit
is assessed through a $\chi^2$-scheme and we use the $\delta\chi^2 =
\chi^2 - \chi^2_{\min}$ statistic to assign confidence values to the
iso-$\chi^2$ contours \citep[e.g.][]{ngc2320-CRdZ, HST-centersIII}.

The main differences with respect to models previously published are
(a) the use of the full non-parametrized LOSVDs \citep[but
see][]{ngc3379-3I, 2001ApJ...550...75B, G3I-03}, (b) the
two-dimensional coverage of the data \citep[but
see][]{2002ApJ...578..787C, sauron-M32}. Most previous studies made
use of long-slit 1D-data at a few position angles
\citep[e.g.][]{m32-axisym}, but the advance of IFU spectrographs (e.g.
\oasis/CFHT, \sauron/WHT, \texttt{FLAMES} and \texttt{VIMOS} at the
Very Large Telescope, etc.) will deliver two-dimensional data for many
objects. A more detailed study of the effect of using such
two-dimensional constraints will be presented in a companion paper
(Cretton \& Emsellem, in preparation).


\subsection{\sauron constraints with original errors}

\begin{figure*}
  \figsc{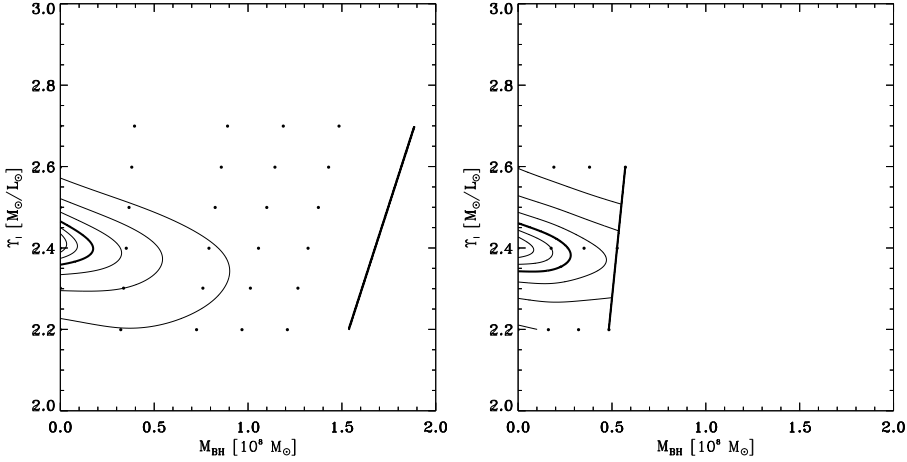} 
  \caption{\emph{Left panel:} $\delta \chi^2$-contours based on
    \sauron data and original (statistic) errors. Each dot represents
    a model run. The \emph{thick contour} corresponds to the 99.7\%
    confidence level. \emph{Right panel:} Same as \emph{left panel},
    but with a larger orbit library ($\times 8$). Differences between
    the two plots are small.}
  \label{fig:chi2_old}
\end{figure*}

We compute a grid of orbit libraries with \Mbh{} ranging from $0.0$ to
$1.82\,10^8$~\Msun{} and \MLI between $2.2$ and $2.6$\footnote{A Jeans
  model based on the previous MGE model (\refsec{sec:mge}) gives $\MLI
  \sim 2.5$ \citep[see e.g.][]{ngc3115-dyn}.}. First, we constrain
these models only with the \sauron dataset and $\delta
\chi^2$-contours are showed in \reffig{fig:chi2_old} (left panel). At
99.7\% confidence level, the allowed \Mbh{} is \emph{smaller} than
$\sim 2\,10^7$~\Msun. This appears to be surprisingly small, given the
low spatial resolution ($\sim 2\farcs1$~FWHM) of the \sauron dataset.
Indeed, in the case of \NGC{3377} with a characteristic central
velocity dispersion of $\sigma \sim 135$~km/s and an assumed distance
of $D = 9.9$~Mpc, this spatial resolution corresponds roughly to the
radius of influence of a $1.6\,10^8$~\Msun{}~BH
\citep{2001bhbg.conf...78D}.

In order to check that the tightness of the $\delta\chi^2$-contours is
not an artifact of our choice of technical implementation, we have
constructed several orbit libraries, changing one parameter of the
library at a time: (a) the length of each orbit integration has been
increased up to 1000 radial periods, (b) the $E$-sampling of the
library has been refined, (c) the individual orbital time-steps have
been decreased, and (d) the size of the orbit library has been
increased by eight-fold, $(N_E, N_{L_{z}}, N_{I_{3}}) = (40,28,14)$
instead of $(20,14,7)$. For the latter model, the resulting
$\delta\chi^2$-contours are shown in the right panel of
\reffig{fig:chi2_old}: although the orbit library was expanded by a
factor of 8, the upper limit for \Mbh{} is still surprisingly small.
In fact, none of the above modifications affected the tightness of the
$\delta\chi^2$-contours in a significant way. We conclude that this
peculiar result is not related to specific technical details of our
modeling method.


\subsection{Departures from axisymmetry}
\label{sec:depart-from-axisymm}

To test further the origin of the tightness of the $\delta
\chi^2$-contours for the \sauron dataset, we constructed fake
constraints drawn from an isotropic distribution function $f(E,L_z)$,
using the \citet{1993MNRAS.262..401H} method (hereafter HQ). The even
part of such a DF is uniquely specified by the mass profile and we
choose the odd part such as to mimic the true observed \sauron
kinematics. A $10^8$~\Msun~BH has been included into the HQ model.
HQ-LOSVDs are then computed in the \sauron setup: they are convolved
with the \sauron PSF and binned into the \sauron spatial elements.
This artificial dataset looks very similar to the \sauron data, but a
dynamical fit gives very different $\delta \chi^2$-contours, allowing
BH masses up to $1.5\,10^8$~\Msun{} at 99.7\% confidence level. For
this fit, we have used ad-hoc error-bars equal to 5\% of the largest
LOSVD value in each pixel. In each \sauron pixel, these errors are
therefore independent of $v_\mathrm{los}$. With such a choice of
errors, we are conservative in the sense that real observed errors are
(on average) three times higher and would therefore induce even wider
$\delta \chi^2$-contours.

By construction, the HQ data correspond to a perfectly axisymmetric
galaxy, and are therefore fully \hbox{[anti-]}bi-symmetric, \ie
$\mathcal{L}_{(x',y')}(v_\mathrm{los}) =
\mathcal{L}_{(-x',y')}(-v_\mathrm{los}) =
\mathcal{L}_{(x',-y')}(v_\mathrm{los})$. In that sense, as mentioned
earlier, the observed \sauron data show noticeable departure from
axisymmetry, and the fit of an axisymmetric model to them will
significantly increase both $\chi^2$ and $\delta\chi^2$.

If one still wishes to model non-axisymmetric data with an
axisymmetric code, there are two possibilities: (a) symmetrize the
data, while keeping the statistical error-bars, (b) increase the
error-bars to encompass the systematic errors (if any) between the
4~quadrants. While the first solution was adopted by G+03, we chose
the second option to keep as much as possible the original spatial
extent of the integral-field data. For the central parts of the
\sauron dataset (roughly $|x'| \le 12\arcsec$ and $|y'| \le
6\arcsec$), for which measures from the 4~quadrants are accessible,
the error-bar including systematics is taken as
$\sigma_{\mathrm{+syst}}^2 = \langle\sigma^2\rangle_{4} +
\mathrm{Var_{4}(\sigma^2)}$, where $\mathrm{Var_{4}}$ is the variance
between the four LOSVDs of the four quadrants and
$\langle\sigma^2\rangle_{4}$ is the mean statistical error-bar of the
four LOSVDs. In the outer parts, the spatial binning and the
incomplete coverage of the \sauron data make the computation more
difficult: we estimate the systematic errors by taking the immediate
neighbor spatial element (after folding all 4~quadrants on one) if
there is no symmetric correspondent in the other quadrants.

As expected, we obtain the same contours using either original or
symmetrized \sauron data when the error-bars including systematics are
used.


\subsection{IFU constraints with errors including systematics}
\label{sec:ifu-constraints}

\begin{figure*}
  \fig{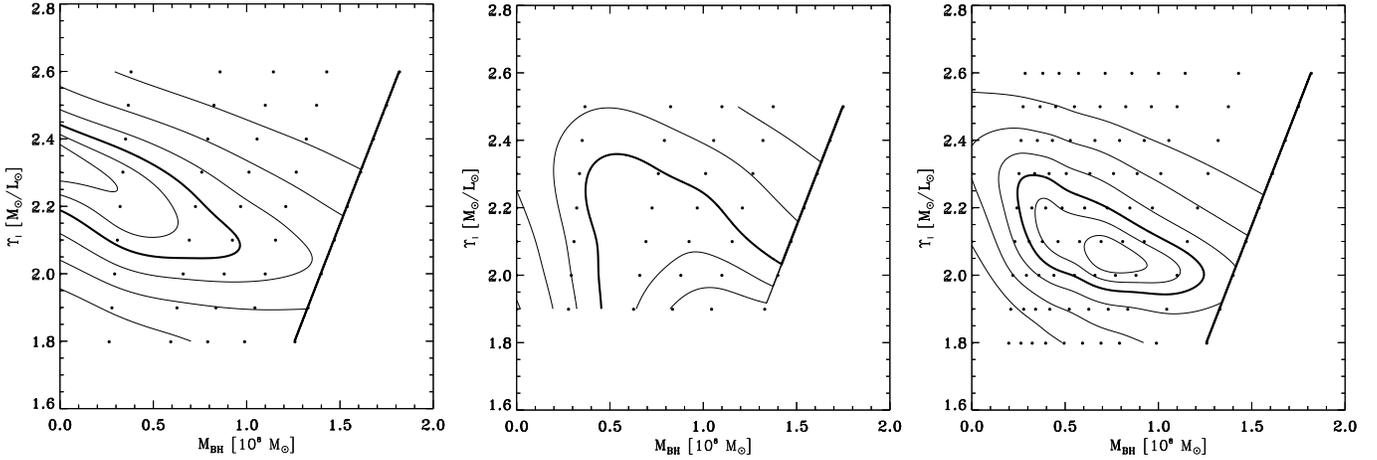}
  \caption{$\delta\chi^2$-contours with errors including systematics. 
    \emph{Left panel:} \sauron dataset only, \emph{middle panel:}
    \oasis dataset only, \emph{right panel:} both \sauron and \oasis
    datasets.}
  \label{fig:chi2_new}
\end{figure*}

\paragraph{\sauron dataset.}
\reffig{fig:chi2_new} (left panel) shows the $\delta\chi^2$-contours
using only the \sauron data with the errors including systematics.
While the constraint on the upper limit of \Mbh{} is noticeably
relaxed (the largest allowed BH mass is now $10^8$~\Msun), \MLI is
strongly restricted to the range $[2.05,2.45]$ (99.7\% confidence
level), due to the large \sauron FoV.

\begin{figure*}
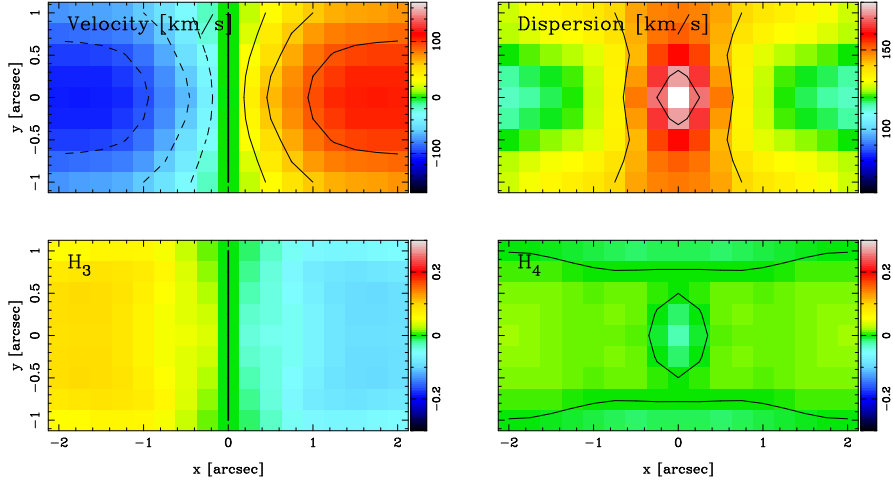

  \figsc{0076-f10} 
  \caption{Best-fit stellar kinematical maps for the \oasis dataset, 
    to be compared with \reffig{fig:oasis-kin}.}
  \label{fig:oasis-fit}
\end{figure*}

\paragraph{\oasis dataset.}
The \oasis data also show signs of non-axisymmetry (see
\reffig{fig:oasis-kin}), so we applied the procedure described in the
previous Sect. to compute systematic error-bars.
\reffig{fig:chi2_new} (middle panel) shows the $\delta
\chi^2$-contours constrained only with the \oasis data (restricted to
the central $4\arcsec \times 2\arcsec$ part, see
\reffig{fig:oasis-fit}). The 99.7\% confidence level is much larger
than in the \sauron case: as could be expected (e.g. G+03) \MLI is
very weakly constrained due to the small extension of the data.
However, a model \emph{without} BH is excluded by the \oasis data at a
99.7\% confidence level.

\paragraph{Combined datasets.}
\reffig{fig:chi2_new} (right panel) shows the results of a combined
fit on both datasets. Since this fit includes all available IFU data,
it presumably provides the best estimate of the BH mass and the
stellar \MLI. To improve such estimates based on a grid interpolation,
we compute additional models corresponding to new values of
$\MLI-\Mbh$. The best fitting model has $\Mbh = 7.0\,10^7$~\Msun{} and
$\MLI = 2.05$ (see the best-fit maps in \reffig{fig:oasis-fit} and
\ref{fig:sauron-fit}, to be compared with the observed maps in
\reffig{fig:oasis-kin} and \ref{fig:sauron-kin}). The minimum and
maximum BH mass allowed by these data (99.7\% confidence level) are
$2.5\,10^7$~\Msun{} and $1.25\,10^8$~\Msun{} respectively.

\begin{figure*}
  \figsc{0076-f11} 
  \caption{Same as \reffig{fig:oasis-fit} for the \sauron dataset, 
    to be compared with \reffig{fig:sauron-kin}.}
  \label{fig:sauron-fit}
\end{figure*}


\subsection{Regularized models}
\label{sec:regularisation}

In previous Sects., we have explored models with no requirements on
the smoothness of the solution. As a result, the orbital weights are
jagged and can strongly vary from one orbit to the next. One can
include regularization constraints in order to obtain smoothly varying
solutions (see e.g. C99). From such regularized solutions smooth
functions characterizing the internal dynamical structure can be
computed (see \reffig{fig:dispersions_BH}). But there is a much more
important reason to derive regularized models. Recently,
\cite{2002astro.ph.10379V} have shown that (unregularized) orbit-based
models do not provide reliable BH~mass estimates, because the
$\chi^2$-contours on which they are based depend strongly on the
number of orbits in the library: if one increases the number of
orbits, the contours get wider and the indeterminacy on the BH grows.
In a forthcoming paper, we have explored this issue with
component-based $f(E,L_z)$ models and have also observed such a
significant widening of the contours when the orbit library is
expanded. Our tests also show that regularization may provide a way to
`stabilize' the $\chi^2$-contours, while increasing the number of
orbits \citetext{for details, see Cretton \& Emsellem, in
  preparation}. Although we do not have yet a clear physical
justification for the regularization scheme, we decided to apply such
a procedure to construct regularized models of \NGC{3377}. We refer
the reader to the companion paper for a discussion of these issues,
outside the scope of this paper.

\begin{figure}
  \fig{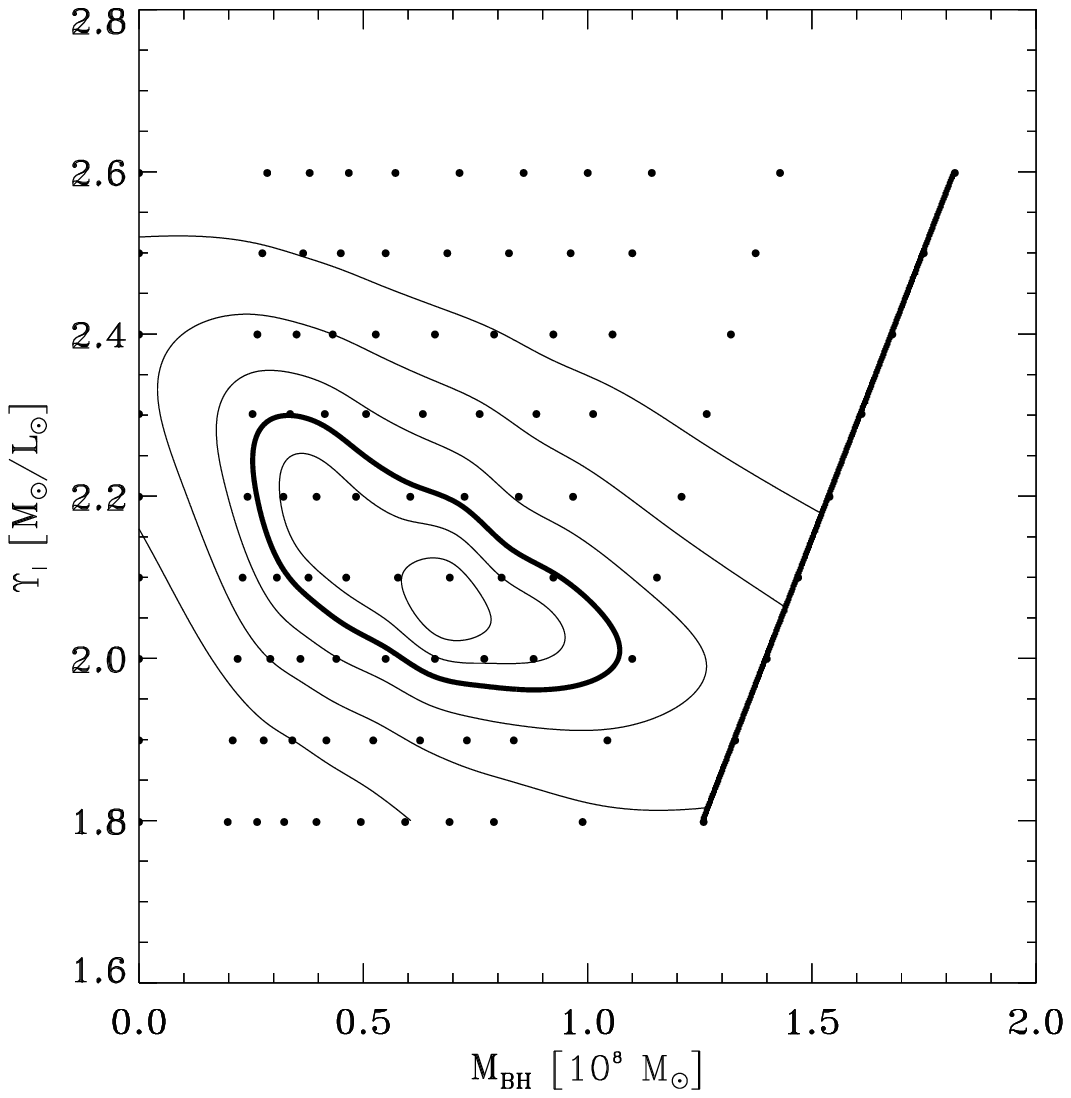}
  \caption{$\delta \chi^2$-contours based on both \sauron and \oasis
    data with errors including systematics \emph{and} regularization
    constraints.}
  \label{fig:chi2_oasis_and_sauron_regul}
\end{figure}

\reffig{fig:chi2_oasis_and_sauron_regul} shows the combined fit,
including regularization constraints. The level of regularization has
been adjusted from a comparison with the HQ models according to the
prescription of C99 (their Fig.~8). The contours are then very similar
to the case without regularization, and the best fitting values are
almost unchanged with respect to the unregularized model with the
exception of the slightly restricted maximum allowed BH mass, which is
now $1.1\,10^8$~\Msun. The values we adopt are thus \Mbh{} in the
range 2--$11\,10^{7}$~\Msun{} (99.7\% confidence level), with a best
fit value of $7\,10^{7}$~\Msun{} for a mass-to-light ratio of $\MLI =
2.1 \pm 0.2$, and an assumed distance of 9.9~Mpc. Note that at
the 68\% confidence level, our model is well constrained with \Mbh{}
between $6$ and $8\,10^{7}$~\Msun.

\subsection{Internal dynamical structure}
\label{sec:internal-dyn}

\begin{figure*}
  \figsc{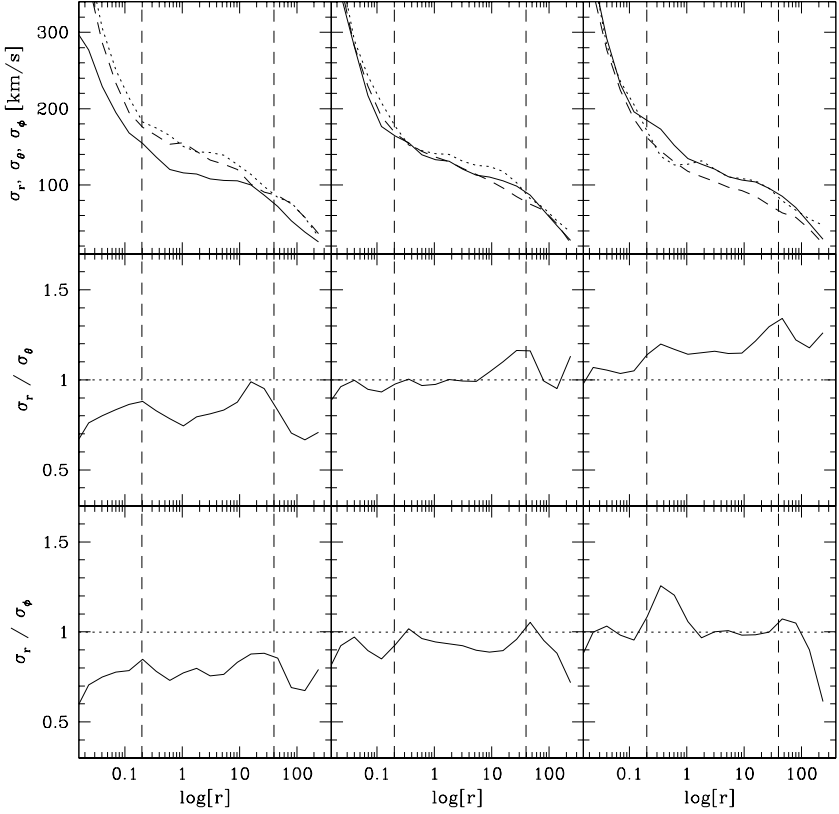} 
  \caption{Internal dynamics of a regularized model with with 
    $\Mbh = 7\,10^7$~\Msun{} and $\MLI = 2.1$ for different angular
    sectors (see text). The \emph{first row} displays the three
    components of the velocity dispersion ellipsoid in km/s:
    $\sigma_{r}$ (\emph{solid line}), $\sigma_{\theta}$ (\emph{dashed
      line}) and $\sigma_{\phi}$ (\emph{dotted line}). The
    \emph{second} and \emph{third rows} show the ratios
    $\sigma_{r}/\sigma_{\theta}$ and $\sigma_{r}/\sigma_{\phi}$.}
  \label{fig:dispersions_BH}
\end{figure*}

To describe the internal dynamical structure of \NGC{3377}, we compute
the velocity dispersion profiles of the best-fit orbit model:
$\sigma_{r}$, $\sigma_{\theta}$ and $\sigma_{\phi}$, where $(r,\theta,\phi)$
are the usual spherical coordinates (see \reffig{fig:dispersions_BH}).
We use the regularized models constrained on both \sauron and \oasis
datasets with errors including systematics (see previous Sect.). In
\reffig{fig:dispersions_BH}, we also plot the ratios
$\sigma_{r}/\sigma_{\theta}$, and $\sigma_{r}/\sigma_{\phi}$ to better
estimate the anisotropy.

The meridional plane $(R,z)$ is divided into a polar grid with seven
angular sectors for $\theta \in [0, \pi/2]$ (see \refsec{sec:schw});
the first sector is close to the symmetry axis, while the last one is
near the equatorial plane. We decide to discard the sector closest to
the symmetry axis (\ie the first one), because only few orbits with
very small $L_z$ can reach it. To reduce the noise, we average sectors
two by two: \#2 with \#3, \#4 with \#5 and finally sectors \#6 with
\#7, in the left, middle and right panels respectively (see
\reffig{fig:dispersions_BH}).

In each panel, the vertical dashed lines indicate the extent of the
kinematic data: it thus delineates the region where the models are
directly constrained. Going from the symmetry axis toward the
equatorial plane, the best-fit model (with $\Mbh = 7\,10^{7}$~\Msun)
shows an increasing radial anisotropy. It is however relatively close
to isotropy, with dispersion ratios $\sigma_{r}/\sigma_{\theta}$ and
$\sigma_{r}/\sigma_{\phi}$ of about 1.2. Except for the peak of
$\sigma_{r}/\sigma_{\phi}$ in the central part of the equatorial plane
profile, the model shows a rather constant anistropy along each
(averaged) sector.

Although our mass model contains two flat Gaussian components
($G'_{5}$ and $G'_{8}$, see \reftab{tab:mge}) with $q'_{j} < 0.24$,
the internal structure of our best-fit model of \NGC{3377} does not
exhibit a strong dynamical signature of a disk like in \objNGC{4342}
\citep[see Fig.~13 and \S7.3 of][]{CvdB99}. Indeed, in the case of
\NGC{3377}, these two flat components never dominate the local mass
density, whereas in \NGC{4342} the outer disk starts to dominate
outside of 4\arcsec, inducing a radial anisotropy.
 
G+03 observed a `peak' in $\sigma_{r}/\sigma_{\theta}$ between
$r=0\farcs1$ and 1\arcsec{} in the equatorial plane (see their
Fig.~13). This ratio reaches 1.5 in their best-fit model, whereas we
observed values below 1.2 in the same radial range. The differences
are certainly due to the significantly different spatial coverage of
the datasets used to constrain the models.

\subsection{Comparison with previous models}
\label{sec:models-comp}

The dynamics of \NGC{3377}, and the presence of a central massive BH,
have been first studied by K+98 with isotropic models, and in greater
details by G+03. The modeling technique (3$I$-\schw) and initial
assumptions (axisymmetry, edge-on, constant mass-to-light ratio) used
by G+03 are similar to the ones we applied on our \NGC{3377} integral
field data. A few important differences should be emphasized though.
Firstly, G+03 constrained the spatial luminosity to be constant on
homothetic ellipsoids, which may cause significant differences in the
resulting dynamical model (and internal structure). However,
considering that isophotes of \NGC{3377} are reasonably well
approximated by ellipses (with constant ellipticity), this is probably
not critical in the present case. Secondly, G+03 included the \fos
apertures in their list of observable constraints. As noted by G+03,
the inclusion of \hst measurements improves the significance of the BH
detection, but does not alter significantly the upper limit of the
mass estimates. The uncertainty on the actual location of the \fos
apertures and its corresponding kinematics (see
\refsec{sec:3377-comp}) furthermore motivated our choice of excluding
this dataset.

G+03 found a BH mass of $\Mbh = 1.0^{+0.9}_{-0.1}\,10^{8}$~\Msun{}
($1\sigma$ range, one degree of freedom, for an assumed distance of
11.2~Mpc). This translates into a mass range for the \Mbh{} of about
$8-17\,10^{7}$~\Msun{} at a distance of 9.9~Mpc. However, as explained
by the authors, the lower limit on \Mbh{} is solely constrained by the
2~nuclear \fos LOSVDs: the fit on their ground-based data --~stellar
$V$ and $\sigma$ measurements from K+98 (see
\refsec{sec:3377-comp})~-- is consistent with the absence of central
BH (their Fig.~7). Our best fit value of $\Mbh = 7\,10^7$~\Msun{} is
just at the edge of the 95\% confidence contours derived by G+03, but
outside their 68\% confidence mass range (see their Fig.~3).

Regarding the mass-to-light ratio, the best-fit value of $\ML[V] =
2.9^{+0.1}_{-0.6}$ at $D = 11.2$~Mpc found by G+03 (99.7\% level)
corresponds to $\MLI \sim \mathbf{2.16} ^{+0.08}_{-0.45}$ at 9.9~Mpc
(for a mean color index $(V-I) = 1.14$,
\citealt{1994A&AS..104..179G}\footnote{Note that although the colour
  profiles presented in \citet{1994A&AS..104..179G} are right, the
  corresponding $\overline{V-I}$ value quoted in Table~5 is incorrect
  (Goudfrooij, private communication).}, and a standard Cousins
$(V-I)_{\sun} = 0.685$, \citealt{1979PASP...91..589B}), consistent
with our best fit value of $\MLI = 2.1$. At the 68\% level, we obtain
a narrow range of allowed \MLI, between 2.02 and 2.12, a significant
improvement compared to the corresponding range of 1.71--2.24 found by
G+03. The main reason for our better mass-to-light ratio constraint
certainly lies in the spatial coverage of our kinematical datasets:
G+03 only disposed of major-axis (long-slit) profiles (with the
addition of the \fos apertures). As shown in \reffig{fig:chi2_new},
the \sauron dataset with its two-dimensional coverage on a relative
large field strongly constrains \ML[]{} (assumed to be constant).

Using the combined \sauron and \oasis datasets, we therefore
significantly improve the constraints on both \Mbh{} and \ML[]. While
our ground-based data spatial resolution is coarser than the \hst
spatial-grade resolution, we believe, given our aforementioned
concerns regarding the uncertainties on critical kinematic
measurements derived from \fos apertures, that our BH mass estimate is
overall more robust than G+03's. In addition, we think that our
mass-to-light ratio estimate is better because of the extended spatial
coverage of our kinematical datasets.


\section{Discussion}
\label{sec:discussions}

The kinematic data used to constrain the dynamical modeling presented
in this paper were exclusively obtained from two integral-field
spectroscopic observations of \NGC{3377}. The full spatial coverage
allows a proper estimate of the respective PSF (see
\refsec{sec:psf-merge}), and a detailed comparison of the datasets,
insuring their internal consistency (see \refsec{sec:3377-comp}).
Furthermore, the spatial complementarity of the datasets~-- \sauron
covers a relatively large FoV while \oasis probes the central parts
with sharper spatial resolution~-- plays a key role in the modeling
process (see \refsec{sec:ifu-constraints}).

Our implementation of the \schw technique has been extended to support
2D-coverage and non-parametrized LOSVD fit, and includes
regularization constraints minimizing the `contours widening' effect
described by \cite{2002astro.ph.10379V}. However, our models still
depend on three major assumptions, namely a given inclination,
axisymmetry, and a constant mass-to-light ratio.

\paragraph{Inclination.}
The galaxy \NGC{3377} is known \emph{not to be} strictly edge-on, as
strongly hinted from the ionized gas distribution \citep[if the gas is
confined in the equatorial plane]{sauron-paper1}. However, \NGC{3377}
is certainly close to edge-on ($i\ga 70^{\circ}$, see
\refsec{sec:deproj}) given its apparent flattening (E5-6). Running a
full grid of dynamical models over the allowed range of inclinations
\citep[as in, e.g.][]{ngc3379-3I, sauron-M32}, might not be relevant
if, as suspected, the assumption of axisymmetry is the most
restrictive one.

\paragraph{Axisymmetry.}
As already mentioned, a $\sim 10^{\circ}$ kinematic misalignment is
observed both in the \oasis and \sauron datasets, in contradiction
with an intrinsic axisymmetric morphology (while the photometry does
not show any position angle twist). This non-axisymmetry could be due
to the presence of an inner bar \citetext{hypothesis supported by the
  gas spiral-like distribution, Emsellem \etal, in preparation}. While
the observed non-axisymmetry weakens the results of our axisymmetric
models, it is not clear whether it is critical for the global
understanding of the intrinsic dynamics of this `nearly oblate' galaxy
\citep{1983ApJ...266...41D} --~contrary to more extreme cases
\citep[e.g. \NGC{4365},][]{sauron-paper3, 2003astro.ph..1070V}~-- and
wether it prevents our $(\Mbh,\ML[])$ best-fit parameters to be taken
into consideration.

\paragraph{Mass-to-light ratio.}
We can compare the dynamical estimate of our \ML[]{} (obtained via
\schw modeling) to the one independently inferred from observed
colours and line strengths. \citet{Idiart03} estimated the metallicity
of \NGC{3377} using broad-band colours, and derived a central and
global $[Fe/H]$ of $+0.0$ and $-0.2$ respectively. Using
\citet{Vazdekis96} stellar population synthesis models with observed
colors $B-V \sim 0.86$ and $V-I \sim 1.14$, this corresponds to a
\ML[V]{} in the (loose) range 2.7--3.7. Our best-fit model has $\MLI =
2.1$, or $\ML[V] = 3.2$, fully consistent with these estimates (but
see discussion in \citealt{maraston98,2003MNRAS.339..897T}).  We also
checked that the \sauron line strength values (\element{H}$\beta$,
\element{Mg}b, \element{Fe}5015, \element{Fe}5270) observed for
\NGC{3377} are within the ranges predicted by the Vazdekis \etal
models. This independent check is important as it links the stellar
populations to the dynamical observables.

Regarding our central BH mass measurement, the only independent
estimate we can provide, beside comparison to previously published
dynamical models (see~\refsec{sec:models-comp}), is by using the
$\Mbh-\sigma$ relation: \citet[]{Merritt01} and
\citet[]{2002ApJ...574..740T} used stellar dispersion (central or
effective) values of 131 and 145~\kms{} respectively, providing $\Mbh
= 1.47 (\pm 0.72)\,10^{7}$ and $\Mbh \simeq 3.7 (\pm
1.7)\,10^{7}$~\Msun, respectively. The range of allowed \Mbh{} derived
from our Schwarzschild models overlap with both estimates, although
our best-fit value is clearly outside these two ranges.

Studying the importance of the effect of triaxiality is outside the
scope of this paper and will be done in the future using generalized
modelling techniques \citetext{such as the one developed by Verolme
  \etal, in preparation}. In that respect, it is noticeable that only
a two-dimensional coverage can properly trace the non-axisymmetry.
This has consequences not only for the specific case of \NGC{3377},
the non-axisymmetry of which was ignored before, but probably for many
other galaxies and related studies.


\section{Conclusions}
\label{sec:concl}

We have presented a unique set of nested integral-field spectroscopic
observations of the E5-6 `disky/boxy' cuspy galaxy \NGC{3377},
obtained with the \oasis and \sauron instruments. Both sets display
regular stellar kinematics, with indications for an inner disk,
coherent with the diskyness of the light distribution up to $\sim
30\arcsec$. However, the IFS observations also reveal a significant
twist of the line of zero velocity of $\sim 10^{\circ}$ up to $r
\approx 10\arcsec$ --~with no corresponding isophotal twist~--
indicative of a moderately triaxial intrinsic shape or of a central
bar.

Based on this complementary and intrinsically coherent dataset, we
have constructed general (three-integral) axisymmetric dynamical
models for this galaxy using the orbit superposition technique of
\schw, non-parametrically fitting the full LOSVDs delivered by the 3D
spectrographs. In order to properly take into account the observed
non-axisymmetry of the kinematic maps, we have added a systematic
component to the Monte-Carlo-computed statistical errors on the LOSVDs
used in the fit. The best-fit model has a BH mass of $\Mbh =
7_{-5}^{+4}\,10^7$~\Msun{} and a mass-to-light ratio $\MLI = 2.1 \pm
0.2$ (99.7\% confidence level) for an assumed distance of $D =
9.9$~Mpc.

The use of fully coherent IFS measurements on the observation side,
and of proper errors and regularization constraints on the modeling
side allowed us to derive this significantly improved constraints on
both \ML[]{} and \Mbh. However, some assumptions used in the modeling
--~the most important presumably being the axisymmetry~-- are in
noticeable contradiction with our IFS observations. Therefore, BH
mass estimates for \NGC{3377} based on axisymmetric models should be
considered with caution.


\begin{acknowledgements}
  YC's research was supported through a European Community Marie Curie
  Fellowship at Leiden Observatory. NC thanks H.-W. Rix for many
  insightful discussions. EE wishes to warmly thank R.~Michard for
  making available the ground-based photometry of \NGC{3377}, and the
  visitor program at ESO, during which part of this work was done. We
  also would like to thank Michele Cappellari, Roger Davies, Harald
  Kuntschner and Tim de Zeeuw for a critical reading of the
  manuscript.
\end{acknowledgements}


\bibliographystyle{aa}
\bibliography{0076}


\appendix


\section{Adaptive quadtree two-dimensional binning}
\label{sec:2d-binning}

In order to increase the $S/N$ in the outer parts of the \sauron
exposure, we used an adaptive spatial binning algorithm. A detailed
account of binning techniques is described in \citet{2D-binning}.

While filtering (\ie `smoothing') spectrographic data is equally easy
for long-slit and integral-field spectrographs, data binning is much
trickier for two-dimensional data, due to the topological constraints
(\ie the proper tiling of the plane). Still, proper binning of
two-dimensional data is necessary when we wish to compare them with
theoretical or numerical models.

The first step is to estimate the flux and mean $S/N$ at each spatial
element (hereafter \emph{spaxel}). In the present case, each spaxel
$i$ of the input cube is associated to the spectrum $S_{i}(\lambda)$,
and one can compute the total flux ${\cal S}_{i} = \int
S_{i}(\lambda)\,\mathrm{d}\lambda$ by integrating along the spectral
dimension, as well as the mean signal-to-noise ratio $\eta_{i} =
\langle S_{i}(\lambda)/\sigma_{i}(\lambda) \rangle_{\lambda}$, where
$\sigma_{i}(\lambda)$ is the chromatic variance as estimated during
the data-reduction process. Furthermore, the real $({\cal S}_{i},
\eta_{i})$-spaxels are embedded in a $2^{n}\times 2^{n}$-grid,
otherwise completed with $(0,0)$-spaxels.

The algorithm is then as follows:
\begin{description}
\item[\textsf{Initialization.}] All the spaxels of the input grid are
  associated in a single bin, with flux ${\cal F}_{1} = \sum_{i}{\cal
    S}_{i}$ and mean signal-to-noise $\zeta_{1} = {\cal
    F}_{1}/\sum_{i}({\cal S}_{i}^{2}/\eta_{i}^{2})$. Hopefully, this
  1-bin has a $S/N$ high enough to be divided further and initialize
  the process.
\item[\textsf{Iteration.}] If $m$-bin $j$ is indeed a bin --~\ie
  contains at least 4~spaxels~-- and has high enough a $S/N$ with
  respect to the target $S/N$, it can be divided further into 4
  subsequent $(m+1)$-bins. Since the initial grid had a power-of-two
  side length, all the child-bins also have a power-of-two side
  length.
\item[\textsf{Conclusion.}] All the bins which are not completely
  covered by real data (in our case, 80\%) or whose $S/N$ is lower
  than a given threshold are then discarded.
\end{description}
This method leads naturally to an \emph{Unbalanced Quadtree spatial
  binning} of the original datacube \citep[e.g.][]{BerEpp-CEG-95}.
Contrarily to the optimal binning presented in \citet{2D-binning},
this quadtree binning does not produce a very homogeneous $S/N$
distribution over the field. However, it provides simple shapes
--~squares~-- for the final bins, and therefore simplify the
comparison with models.

\begin{figure}
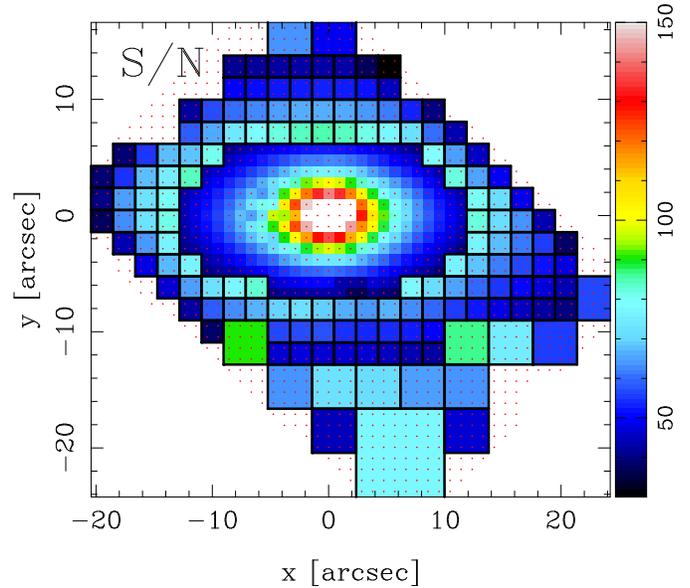

  \fig{0076-f14}
  \caption{Unbalanced Quadtree spatial binning of the \sauron datacube
    resulting from the method described in the text. The \emph{dots}
    show the location of the initial \sauron spaxels, and the
    \emph{squares} display the boundaries of the multi-spaxel bins.}
  \label{fig:binning}
\end{figure}

For the specific case of the \sauron \NGC{3377} data, we have used a
target $S/N$ of~50 and a minimum $S/N$ of~30. This resulted into
475~independent binned spectra as shown in \reffig{fig:binning}.

\end{document}